\documentclass{optica-article}

\journal{opticajournal} 

\articletype{Research Article}

\usepackage{graphicx} 
\usepackage{float} 
\usepackage{wrapfig} 
\usepackage[labelfont=bf]{caption} 
\usepackage{caption, subcaption} 

\colorlet{mycolor1}{orange}                  
\colorlet{mycolor2}{blue!90!black}           
\colorlet{mycolor3}{blue!14}                 
\colorlet{mycolor4}{blue!40}                 

\usepackage{tikz,tikz-3dplot}
\usetikzlibrary{arrows.meta,bending,matrix,positioning,patterns,fit,calc,arrows,shapes,angles,quotes,spy,intersections,decorations.markings,math,3d,perspective}
\tikzstyle{startstop} = [ellipse, thick, minimum width=3cm, minimum height=0.5cm, text centered, draw=black, fill=mycolor1, align=center]
\tikzstyle{process} = [rectangle, thick, rounded corners, minimum width=3cm, minimum height=0.5cm, text centered, draw=black, fill=mycolor4, align=center]
\tikzstyle{decision} = [ellipse, thick, minimum width=3cm, minimum height=0.5cm, text centered, draw=black, fill=mycolor4]
\tikzstyle{arrow} = [arrows = {-Stealth[length=8pt, inset=2pt,round]}]

\usepackage{bm} 
\usepackage{hyperref} 
\usepackage{lineno} 

\begin{document}

\pagestyle{plain}  
\pagenumbering{arabic}  

\title{An inverse design method for generalized zero-étendue sources and two targets}
\author{P. A. Braam,\authormark{1,*} J. H. M. ten Thije Boonkkamp,\authormark{1} M. J. H. Anthonissen,\authormark{1} K. Mitra,\authormark{1} L. Kusch,\authormark{1} and W. L. IJzerman\authormark{1,2}}
\address{\authormark{1}CASA, Department of Mathematics and Computer Science, Eindhoven University of Technology, \mbox{P.O. Box 513,} 5600 MB Eindhoven, The Netherlands\\
\authormark{2}Signify Research, High Tech Campus 7, 5656 AE Eindhoven, The Netherlands}
\email{\authormark{*}p.a.braam@tue.nl}

\begin{abstract*} 
We present an inverse method to compute freeform optical surfaces that transform a light distribution, parameterized by two source planes, into two separate target distributions. The surfaces can be reflectors or lenses, and control both the spatial and directional source and target coordinates of light rays. From energy conservation we derive Jacobian equations for optical mappings, and the optical path length provides generating functions for the optical surfaces. A three-stage least-squares algorithm numerically solves the resulting equations. We present examples with complex source and target distributions.
\end{abstract*}

\section{Introduction}
In non-imaging optics, optical surfaces are used to transform light from a given source emittance to a desired target irradiance distribution. While these surfaces may exhibit geometric symmetries in certain cases, their shapes are in general non-symmetric, often referred to as \textit{freeform}. The goal in freeform optical design is to determine the shapes and positions of these optical surfaces, which can be either reflector (mirror) or lens surfaces.

Freeform optical design methods can be broadly categorized as \textit{forward} or \textit{inverse} methods. In forward methods, a specified emittance distribution at the source and a given optical system are used to sample rays randomly, often using Monte-Carlo ray tracing techniques, to determine the target irradiance distribution \cite{Li2018,Carmela2016,Carmela2017,Carmela2021,Carmela2018}. The optical system is then iteratively adjusted to improve this irradiance. Although achieving high precision can be a slow process, recently developed methods have shown ways to improve the accuracy and to reduce computation time \cite{Wang2022,Wang2023,Tang2024,Willem2024}. 

In contrast, inverse methods compute the optical surfaces from the source and target distributions by applying principles of geometrical optics and energy conservation. There are a variety of approaches to solve the resulting equations. We mention a few and refer to Romijn \textit{et al.}\ \cite{Lotte2021} for a comprehensive overview. One approach is to solve a partial differential equation of Monge-Ampère type by discretization. To this end, Kawecki \textit{et al.\ }used a finite element method \cite{Kawecki2018}, both Benamou \textit{et al.}\ and Froese \textit{et al.}\ used finite differences \cite{Benamou2010,Benamou2014,Benamou2015,Benamou2019,Benamou2022,Froese2012}, Brix \textit{et al.}\ used a collocation method \cite{Brix2015,Brix2015_} and Hacking \textit{et al.}\ used a neural network \cite{Roel2025}. Another approach, proposed by Doskolovich \textit{et al.}, applies an optimization method using theory on optimal transport \cite{Doskolovich2019}. Caffarelli and Oliker introduced an approach based on geometric constructions \cite{Caffarelli2008}, which has since been used by others to construct optical surfaces \cite{Kitawaga2019,Merigot2021,Gallouet2021}.

This article focuses on an approach first introduced by Caboussat \textit{et al.}, which solves the standard Monge-Ampère problem with Dirichlet boundary conditions using a least-squares method \cite{Caboussat2013}. Prins \textit{et al.}\ adapted this approach by including the more suitable \textit{transport boundary condition} \cite{Corien2015}, initially introduced by Benamou \textit{et al.}\ \cite{Benamou2014}, ensuring that all light rays from the source reach the target. Yadav \textit{et al.}\ extended this approach by solving a generalized Monge-Ampère equation for designing systems with multiple optical surfaces \cite{Nitin2019,Nitin2019_,Nitin2019__}. Subsequently, Romijn \textit{et al.}\ adjusted the method further to solve generalized Jacobian equations \cite{Lotte2019,Lotte2020,Lotte2021_,Lotte2021__}, resulting in the \textit{generating least-squares method}. This method has since been applied to model various optical systems \cite{Teun2021,Pieter2024,Jan2024}.

Two optical systems that convert a parallel-source emittance to a parallel-target irradiance are the parallel-to-parallel reflector system which uses two reflectors and the parallel-to-parallel lens system which uses two lens surfaces \cite{Martijn2021}. To compute the surfaces in the reflector system, Oliker \textit{et al.}\ used the supporting quadric method \cite{Oliker2007} and both Bösel \textit{et al.}\ and Feng \textit{et al.}\ used ray-mapping approaches \cite{Bosel2018,Feng2017}. To compute the surfaces in the lens system, Oliker \textit{et al.}\ again applied the supporting quadric method \cite{Oliker2018}, Bösel \textit{et al.}\ and Feng \textit{et al.}\ used ray-mapping methods \cite{Bosel2016,Bosel2018,Feng2015,Feng2013,Feng2021}, Doskolovich \textit{et al.}\ reduced the system to a linear assignment problem \cite{Doskolovich2018} and Zhang \textit{et al.}\ solved a nonlinear boundary value problem using Newton's method. Both systems were also studied by Yadav \textit{et al.}\ \cite{Nitin2018_,Nitin2019,Nitin2019__} and Van Roosmalen \textit{et al.} \cite{Teun2024}, whose work this article builds upon.

For the parallel-to-parallel systems, we denote the distance from the source $\mathcal{S}$ to the first optical surface by $u_1(\bm{x})$ for coordinates $\bm{x}$ at $\mathcal{S}$ and from the second optical surface to the target $\mathcal{T}$ by $u_2(\bm{y})$ for coordinates $\bm{y}$ at $\mathcal{T}$. The generating least-squares method is based on this formulation and uses a \textit{generating function} $G$ of the form
\begin{align}\label{eq:introduction_G}
    u_1(\bm{x})=G(\bm{x},\bm{y},u_2(\bm{y});V),
\end{align}
where $V$ is the optical path length. In the parallel-to-parallel systems, $V$ is constant. However, when the direction of light rays vary at the source and target, the optical path length is a function of the source and target coordinates, i.e., $V=V(\bm{x},\bm{y})$. As a result, Eq.~(\ref{eq:introduction_G}) can be generalized to
\begin{align}\label{eq:general_generating_function}
    u_1(\bm{x})=G(\bm{x},\bm{y},u_2(\bm{y});V(\bm{x},\bm{y})).
\end{align}
We will derive a method for finding $V(\bm{x},\bm{y})$, which can be used in the generating function $G$ to determine the optical surfaces.

\begin{figure}[!ht]
  \centering
\begin{subfigure}[b]{0.43\textwidth}
    \centering
    \begin{center}
\begin{tikzpicture}[scale=0.55]
    \draw [line width=0.7mm] (0,-1) -- (5,-1); 
    \draw [line width=0.7mm] (0,0) -- (5,0); 
    \draw [line width=0.7mm] (6.2,5) -- (11.2,5); 
    \draw [line width=0.7mm] (6.2,6) -- (11.2,6); 
    \draw [line width=0.7mm] (5,4.52) arc (115:135:17.8cm); 
    \draw [line width=0.7mm] (6.2,0.5) arc (295:315:17.7cm); 
    \draw [dotted, line width=0.25mm] (0,-1) -- (0,1) -- (6.2,0.6) -- (6.2,6); 
    \draw [dotted, line width=0.25mm] (1,-1) -- (1.25,2) -- (7.7,1.4) -- (8.5,6); 
    \draw [dotted, line width=0.25mm] (2.2,-1) -- (2.5,3.1) -- (9,2.2) -- (9.5,6); 
    \draw [dotted, line width=0.25mm] (3.6,-1) -- (4,4) -- (10,3.05) -- (10.5,6); 
    \draw [dotted, line width=0.25mm] (5,-1) -- (5,4.5) -- (11.18,4) -- (11.18,6); 
    \draw [arrows = {-Stealth[length=8pt, inset=2pt,round]}] (3.5,0.77) -- (3.6,0.765); 
    \draw [arrows = {-Stealth[length=8pt, inset=2pt,round]}] (4.5,1.69) -- (4.6,1.685); 
    \draw [arrows = {-Stealth[length=8pt, inset=2pt,round]}] (5.5,2.685) -- (5.6,2.675); 
    \draw [arrows = {-Stealth[length=8pt, inset=2pt,round]}] (6.6,3.591) -- (6.75,3.57); 
    \draw [arrows = {-Stealth[length=8pt, inset=2pt,round]}] (7.5,4.29) -- (7.6,4.284); 
    \draw [line width=0.25mm, dashed] (5,-1) -- (11.2,-1);
    \draw [line width=0.25mm, dashed] (5,0) -- (11.2,0);
    \draw [line width=0.25mm, dashed] (0,5) -- (8,5);
    \draw [line width=0.25mm, dashed] (0,6) -- (8,6);
    \draw [orange, dashed, line width=0.3mm, dash pattern=on 2pt off 1pt] plot[smooth] coordinates {(0,-0.65) (1,-0.75) (2,-0.75) (3.3,-0.8) (5,-0.85)};
    \draw [blue, dashed, line width=0.3mm, dash pattern=on 2pt off 1pt] plot[smooth] coordinates {(0,0.3) (1,0.3) (2,0.3) (3.3,0.2) (5,0.3)};
    \draw [green, dashed, line width=0.3mm, dash pattern=on 2pt off 1pt] plot[smooth] coordinates {(6.2,5.1) (7.2,5.1) (8.2,5.2) (9.2,5.25) (10.2,5.25) (11.2,5.25)};
    \draw [red, dashed, line width=0.3mm, dash pattern=on 2pt off 1pt] plot[smooth] coordinates {(6.2,6.1) (7.2,6.1) (8.2,6.15) (9.2,6.2) (10.2,6.3) (11.2,6.4)};
    \node at (0.6,-0.4) [orange]{$f_1$};
    \node at (0.6,0.65) [blue]{$f_2$};
    \node at (7,5.5) [green]{$g_1$};
    \node at (7,6.5) [red]{$g_2$};
    \node at (1,2.7) {$\mathcal{R}_1$};
    \node at (10.65,2.65) {$\mathcal{R}_2$};
    \node at (-0.5,-1) {$\mathcal{S}_1$};
    \node at (-0.5,0) {$\mathcal{S}_2$};
    \node at (11.7,5) {$\mathcal{T}_1$};
    \node at (11.7,6) {$\mathcal{T}_2$};
\end{tikzpicture}
\end{center}
    \caption{Reflector system.}
    \label{fig:two-reflectors-sketch}
\end{subfigure}
\hfill
\begin{subfigure}[b]{0.43\textwidth}
    \begin{center}
\begin{tikzpicture}[scale=0.55]
    \path[fill=gray!30] 
        (0,2) arc (240:265:12.2cm) -- (6,3) arc (60:85:12.2cm) -- cycle;
    \draw [line width=0.7mm] (0,-1) -- (5,-1); 
    \draw [line width=0.7mm] (0,0) -- (5,0); 
    \draw [line width=0.7mm] (1,5) -- (6,5); 
    \draw [line width=0.7mm] (1,6) -- (6,6); 
    \draw [line width=0.7mm] (0,2) arc (240:265:12.2cm); 
    \draw [line width=0.7mm] (6,3) arc (60:85:12.2cm); 
    \draw [dotted, line width=0.25mm] (0,-1) -- (0,2) -- (1,4.5) -- (1,6); 
    \draw [dotted, line width=0.25mm] (1.1,-1) -- (0.55,1.6) -- (1.4,4.6) -- (2.1,6); 
    \draw [dotted, line width=0.25mm] (2,-1) -- (1.3,1.4) -- (2.3,4.6) -- (3,6); 
    \draw [dotted, line width=0.25mm] (2.5,-0.9) -- (2.5,1) -- (3.4,4.1) -- (3.5,6); 
    \draw [dotted, line width=0.25mm] (3,-1) -- (3.7,0.7) -- (4.3,3.6) -- (4,6); 
    \draw [dotted, line width=0.25mm] (3.9,-1) -- (4.55,0.6) -- (5.2,3.3) -- (4.9,6); 
    \draw [dotted, line width=0.25mm] (5,-1) -- (5,0.5) -- (6,3.1) -- (6,6); 
    \draw [arrows = {-Stealth[length=8pt, inset=2pt,round]}] (0.65,3.59) -- (0.66,3.61); 
    \draw [arrows = {-Stealth[length=8pt, inset=2pt,round]}] (1.06,3.39) -- (1.066,3.41); 
    \draw [arrows = {-Stealth[length=8pt, inset=2pt,round]}] (1.85,3.09) -- (1.857,3.11); 
    \draw [arrows = {-Stealth[length=8pt, inset=2pt,round]}] (3.00,2.69) -- (3.005,2.71); 
    \draw [arrows = {-Stealth[length=8pt, inset=2pt,round]}] (4.05,2.39) -- (4.054,2.41); 
    \draw [arrows = {-Stealth[length=8pt, inset=2pt,round]}] (4.95,2.19) -- (4.955,2.21); 
    \draw [arrows = {-Stealth[length=8pt, inset=2pt,round]}] (5.58,1.99) -- (5.587,2.01); 
    \draw [line width=0.25mm, dashed] (5,-1) -- (6,-1);
    \draw [line width=0.25mm, dashed] (5,0) -- (6,0);
    \draw [line width=0.25mm, dashed] (0,5) -- (1,5);
    \draw [line width=0.25mm, dashed] (0,6) -- (1,6);
    \draw [orange, dashed, line width=0.3mm, dash pattern=on 2pt off 1pt] plot[smooth] coordinates {(0,-0.8) (1.33,-0.75) (2.5,-0.5) (3.66,-0.75) (5,-0.8)};
    \draw [blue, dashed, line width=0.3mm, dash pattern=on 2pt off 1pt] plot[smooth] coordinates {(0,0.2) (1.33,0.25) (2.5,0.35) (3.66,0.25) (5,0.2)};
    \draw [green, dashed, line width=0.3mm, dash pattern=on 2pt off 1pt] plot[smooth] coordinates {(1,5.2) (2.33,5.25) (3.5,5.35) (4.66,5.25) (6,5.2)};
    \draw [red, dashed, line width=0.3mm, dash pattern=on 2pt off 1pt] plot[smooth] coordinates {(1,6.2) (2.33,6.25) (3.5,6.5) (4.66,6.25) (6,6.2)};
    \node at (0.4,-0.4) [orange]{$f_1$};
    \node at (0.4,0.65) [blue]{$f_2$};
    \node at (1.4,5.5) [green]{$g_1$};
    \node at (1.4,6.5) [red]{$g_2$};
    \node at (-0.5,2) {$\mathcal{L}_1$};
    \node at (6.5,3) {$\mathcal{L}_2$};
    \node at (-0.5,-1) {$\mathcal{S}_1$};
    \node at (-0.5,0) {$\mathcal{S}_2$};
    \node at (6.5,5) {$\mathcal{T}_1$};
    \node at (6.5,6) {$\mathcal{T}_2$};
\end{tikzpicture}
\end{center}
\caption{Lens system.}
\label{fig:two-lenses-sketch}
\end{subfigure}
\caption{Sketch of a reflector system with two freeform reflectors $\mathcal{R}_1$ and $\mathcal{R}_2$, and a lens system with two freeform lens surfaces $\mathcal{L}_1$ and $\mathcal{L}_2$, where light rays originate from sources $\mathcal{S}_1$ with distribution $f_1$ and $\mathcal{S}_2$ with distribution $f_2$ and are directed to targets $\mathcal{T}_1$ with distribution $g_1$ and $\mathcal{T}_2$ with distribution $g_2$.}
\end{figure}
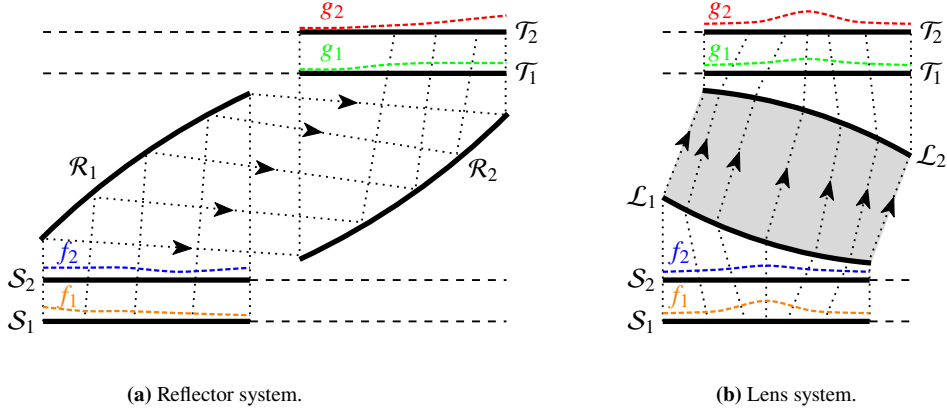

In this paper, we discuss two optical systems: the reflector system given in Fig.~\ref{fig:two-reflectors-sketch} and the lens system given in Fig.~\ref{fig:two-lenses-sketch}. In both systems, the spatial positions and directions of the light rays are fixed by choosing light distributions $f_1$ and $f_2$ on sources $\mathcal{S}_1$ and $\mathcal{S}_2$ indicated by the blue and orange dashed lines. The goal is to determine the two freeform optical surfaces in each system such that the light rays create the desired spatial positions and directions specified by light distributions $g_1$ and $g_2$ on targets $\mathcal{T}_1$ and $\mathcal{T}_2$, indicated by the green and red dashed lines. Here, we assume that each light ray originating from sources $\mathcal{S}_1$ and $\mathcal{S}_2$ hits the optical surfaces in a unique point. 

To determine the shapes of the optical surfaces in both systems, we present a mathematical inverse model. This model is the same for both the reflector and lens system, apart from a different generating function $G$. We use conservation of energy to determine the optical mappings and the optical path length to construct the generating functions. A least-squares algorithm solves the model numerically.

We can describe zero-étendue sources in a general way using light distributions on two source planes. A particular case of the reflector system where light distributions $f_1$ and $f_2$ are the same and thereby form a parallel source can be found in \cite{Braam2025}. A specific case of the lens system where the light distribution $f_1$ is only non-zero in a single point, hereby forming a point source, can be found in \cite{Feng2017,Feng2021}. However, to the best of our knowledge, there is no existing literature on optical systems with either two reflectors or two lens surfaces that prescribe both the positions and directions of light rays at the source and target in a general sense, for instance, by using two source planes and two target planes as we do here.

This paper is structured as follows. In Sec.~\ref{sec:Formulation} we present the model for the two optical systems using generated functions, the optical path length and energy conservation. In Sec.~\ref{sec:Algorithm} we provide an overview of the generating least-squares algorithm. We present results in Sec.~\ref{sec:results} and formulate our conclusions in Sec.~\ref{sec:conclusion}.

\section{Mathematical model}\label{sec:Formulation}
In this section, we first formulate a mathematical model for the reflector and lens system by deriving generating functions. Next, we use energy conservation to find generated Jacobian equations for the optical mappings, which are used in combination with the generating functions to determine the shapes of the optical surfaces. We end this section by discussing how the second target can be replaced by a far-field target.

In this paper, we denote two-dimensional and three-dimensional vectors by a bold symbol, underline them when they are three-dimensional, and add a hat to indicate unit length, e.g., $\underline{\hat{\bm{s}}}$. In addition, $|\cdot|$ denotes the 2-norm.

\subsection{Geometrical description of the reflector system}\label{sec:reflector_description}
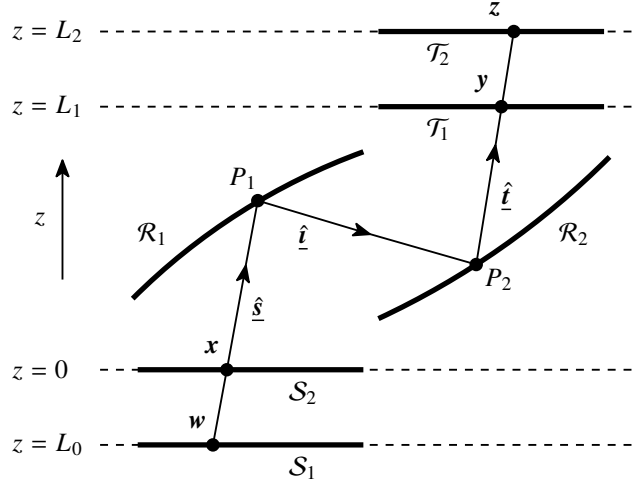
\begin{figure}[!ht]
\begin{center}
\begin{tikzpicture}[scale=1.0]
    \draw [line width=0.7mm] (0,-1) -- (3,-1); 
    \draw [line width=0.7mm] (0,0) -- (3,0); 
    \draw [line width=0.7mm] (3.2,3.5) -- (6.2,3.5); 
    \draw [line width=0.7mm] (3.2,4.5) -- (6.2,4.5); 
    \draw [line width=0.7mm] (3,2.9) arc (110:135:8.4cm); 
    \draw [line width=0.7mm] (3.2,0.68) arc (295:315:10.8cm); 
    \draw [line width=0.25mm, dashed] (-0.5,-1) -- (6.6,-1); 
    \draw [line width=0.25mm, dashed] (-0.5,0) -- (6.6,0); 
    \draw [line width=0.25mm, dashed] (-0.5,3.5) -- (6.6,3.5); 
    \draw [line width=0.25mm, dashed] (-0.5,4.5) -- (6.6,4.5); 
    \draw [line width=0.25mm, arrows = {-Stealth[length=8pt, inset=2pt,round]}] (-1,1.2) -- (-1,2.8); 
    \draw [line width=0.25mm, arrows = {-Stealth[length=8pt, inset=2pt,round]}] (1.185,0) -- (1.44,1.4); 
    \draw [line width=0.25mm, arrows = {-Stealth[length=8pt, inset=2pt,round]}] (1.5965,2.25) -- (3.1,1.8); 
    \draw [line width=0.25mm, arrows = {-Stealth[length=8pt, inset=2pt,round]}] (4.5,1.4) -- (4.7585,3); 
    \draw [line width=0.25mm] (1,-1) -- (1.185,0); 
    \draw [line width=0.25mm] (1.42,1.3) -- (1.5965,2.25); 
    \draw [line width=0.25mm] (3,1.83) -- (4.5,1.4); 
    \draw [line width=0.25mm] (4.751,2.95) -- (5,4.5); 
    \node at (1,-1)[circle,fill,inner sep=1.8pt]{}; 
    \node at (1.185,0)[circle,fill,inner sep=1.8pt]{}; 
    \node at (1.5965,2.25)[circle,fill,inner sep=1.8pt]{}; 
    \node at (4.5,1.4)[circle,fill,inner sep=1.8pt]{}; 
    \node at (4.835,3.5)[circle,fill,inner sep=1.8pt]{}; 
    \node at (5,4.5)[circle,fill,inner sep=1.8pt]{}; 
    \node at (-1.3,2) {$z$};
    \node at (-1.2,-1) {$z=L_0$};
    \node at (-1.3,0) {$z=0$};
    \node at (-1.2,3.5) {$z=L_1$};
    \node at (-1.2,4.5) {$z=L_2$};
    \node at (0.8,-0.7) {$\bm{w}$};
    \node at (1.0,0.3) {$\bm{x}$};
    \node at (1.4,2.55) {$P_1$};
    \node at (4.8,1.2) {$P_2$};
    \node at (4.585,3.8) {$\bm{y}$};
    \node at (4.75,4.8) {$\bm{z}$};
    \node at (1.6,0.8) {$\underline{\hat{\bm{s}}}$};
    \node at (2.2,1.75) {$\underline{\hat{\bm{\imath}}}$};
    \node at (4.9,2.3) {$\underline{\hat{\bm{t}}}$};
    \node at (0.2,1.8) {$\mathcal{R}_1$};
    \node at (5.8,1.8) {$\mathcal{R}_2$};
    \node at (2.2,-1.3) {$\mathcal{S}_1$};
    \node at (2.2,-0.3) {$\mathcal{S}_2$};
    \node at (4,3.2) {$\mathcal{T}_1$};
    \node at (4,4.2) {$\mathcal{T}_2$};
\end{tikzpicture}
\end{center}
\caption{Sketch of the reflector system.}
\label{fig:2S2T_reflector_system}
\end{figure}

We consider the reflector system illustrated in Fig.~\ref{fig:2S2T_reflector_system}, where a light ray propagates from $(\bm{w},L_0)$ with $L_0<0$ on the first source plane $\mathcal{S}_1$ to $(\bm{x},0)$ on the second source plane $\mathcal{S}_2$ with direction $\underline{\hat{\bm{s}}}=\underline{\hat{\bm{s}}}(\bm{x})$. It then hits the first reflector $\mathcal{R}_1$ at point $P_1$ and reflects in direction $\underline{\hat{\bm{\imath}}}=\underline{\hat{\bm{\imath}}}(\bm{x},\bm{y})$. Next, it hits the second reflector $\mathcal{R}_2$ at point $P_2$ and reflects again with direction $\underline{\hat{\bm{t}}}=\underline{\hat{\bm{t}}}(\bm{y})$. The ray then reaches $(\bm{y},L_1)$ on the first target plane $\mathcal{T}_1$, maintains its direction $\underline{\hat{\bm{t}}}$ and finally reaches $(\bm{z},L_2)$ with $L_2>L_1$ on the second target plane $\mathcal{T}_2$. In this optical system, $u_1=u_1(\bm{x})$ is the distance from $(\bm{x},0)$ to $P_1$, $d=d(\bm{x},\bm{y})$ is the distance from $P_1$ to $P_2$ and $u_2=u_2(\bm{y})$ is the distance from $P_2$ to $(\bm{y},L_1)$. The reflectors can be parameterized as
\begin{align}\label{eq:parameterizations_reflectors}
    \mathcal{R}_1: \underline{\bm{r}}_1=\underline{\bm{r}}_1(\bm{x})=\underline{\bm{x}}+u_1(\bm{x})\underline{\hat{\bm{s}}}\in\mathbb{R}^3,
    &&
    \mathcal{R}_2: \underline{\bm{r}}_2=\underline{\bm{r}}_2(\bm{y})=\underline{\bm{y}}-u_2(\bm{y})\underline{\hat{\bm{t}}}\in\mathbb{R}^3,
\end{align}
where $\underline{\bm{x}}=(\bm{x},0)^T$, $\underline{\bm{y}}=(\bm{y},L_1)^T$, and where the source vector $\underline{\hat{\bm{s}}}$ and target vector $\underline{\hat{\bm{t}}}$ are given by
\begin{align}\label{eq:vectors_s_and_t}
    \underline{\hat{\bm{s}}}=\begin{pmatrix}
        \bm{p}_\mathrm{s}\\
        s_3
    \end{pmatrix}
    =
    \frac{1}{\sqrt{|\bm{x}-\bm{w}|^2+L_0^2}}
    \begin{pmatrix}
        \bm{x}-\bm{w}\\
        -L_0
    \end{pmatrix},
    &&
    \underline{\hat{\bm{t}}}=\begin{pmatrix}
        \bm{p}_\mathrm{t}\\
        t_3
    \end{pmatrix}
    =
    \frac{1}{\sqrt{|\bm{z}-\bm{y}|^2+(L_2-L_1)^2}}
    \begin{pmatrix}
        \bm{z}-\bm{y}\\
        L_2-L_1
    \end{pmatrix}.
\end{align}
Since the source vector $\underline{\hat{\bm{s}}}$ is a function of the source coordinates, the light source can be referred to as zero-étendue \cite{Chaves2016}.

Next, we derive the generating function $G_\text{R}$ of the reflector system and its corresponding inverse $H_\text{R}$. To this end, we consider the optical path length $V(\bm{x},\bm{y})$ from $\mathcal{S}_2$ to $\mathcal{T}_1$, which can be written as
\begin{align}\label{eq:2S2T_OPL_V_reflector}
    V=V(\bm{x},\bm{y})=u_1(\bm{x})+d(\bm{x},\bm{y})+u_2(\bm{y}), &&
    d=d(\bm{x},\bm{y})=\big|\;\underline{\bm{y}}-\underline{\bm{x}}-u_1\underline{\hat{\bm{s}}}-u_2\underline{\hat{\bm{t}}}\;\big|.
\end{align}
By rewriting the first equation as $d=V-u_1-u_2$ and squaring both sides, we obtain
\begin{align}\label{eq:distance_d_reflector_squared}
    d^2=V^2+u_1^2+u_2^2+2u_1 u_2-2V(u_1+u_2).
\end{align}
If we square both sides of the expression for $d$ in Eq.~(\ref{eq:2S2T_OPL_V_reflector}) and set it equal to $d^2$ in Eq.~(\ref{eq:distance_d_reflector_squared}), we find the equation
\begin{align}\label{eq:V1_squared}
    0&=|\underline{\bm{y}}-\underline{\bm{x}}|^2-V^2+2u_1\Big(V-(\underline{\bm{y}}-\underline{\bm{x}})\bm{\cdot}\underline{\hat{\bm{s}}}\Big)+2u_2\Big(V-(\underline{\bm{y}}-\underline{\bm{x}})\bm{\cdot}\underline{\hat{\bm{t}}}\Big)-2u_1u_2\Big(1-\underline{\hat{\bm{s}}}\bm{\cdot}\underline{\hat{\bm{t}}}\Big).
\end{align}
From Eq.~(\ref{eq:V1_squared}) we derive that $u_1(\bm{x})=G_\text{R}(\bm{x},\bm{y},u_2(\bm{y});V(\bm{x},\bm{y}))$ with generating function $G_\text{R}(\bm{x},\bm{y},w;V)$ defined as
\begin{subequations}
\label{eq:generating_functions_reflector}
\begin{align}
    G_\text{R}=G_\text{R}(\bm{x},\bm{y},w;V)&=\frac{a_1(\bm{x},\bm{y};V)-a_2(\bm{x},\bm{y};V)w}{a_3(\bm{x},\bm{y};V)-a_4(\bm{x},\bm{y})w},\label{eq:2T_generating_function_G}
\end{align}
and from Eq.~(\ref{eq:V1_squared}) we also derive that $u_2(\bm{y})=H_\text{R}(\bm{x},\bm{y},u_1(\bm{x});V(\bm{x},\bm{y}))$ with inverse generating function $H_R(\bm{x},\bm{y},w;V)$ defined as
\begin{align}
    H_\text{R}=H_\text{R}(\bm{x},\bm{y},w;V)&=\frac{a_1(\bm{x},\bm{y};V)-a_3(\bm{x},\bm{y};V)w}{a_2(\bm{x},\bm{y};V)-a_4(\bm{x},\bm{y})w},\label{eq:2T_inverse_generating_function_H}
\end{align}
with auxiliary variables
\begin{align}
    a_1=a_1(\bm{x},\bm{y};V)&=\tfrac{1}{2}(V^2-|\underline{\bm{y}}-\underline{\bm{x}}|^2),\\
    a_2=a_2(\bm{x},\bm{y};V)&=V-(\underline{\bm{y}}-\underline{\bm{x}})\bm{\cdot}\underline{\hat{\bm{t}}},\\
    a_3=a_3(\bm{x},\bm{y};V)&=V-(\underline{\bm{y}}-\underline{\bm{x}})\bm{\cdot}\underline{\hat{\bm{s}}},\\
    a_4=a_4(\bm{x},\bm{y})&=1-\underline{\hat{\bm{s}}}\bm{\cdot}\underline{\hat{\bm{t}}}.
\end{align}
\end{subequations}

\subsection{Geometrical description of the lens system}\label{sec:lens_description}
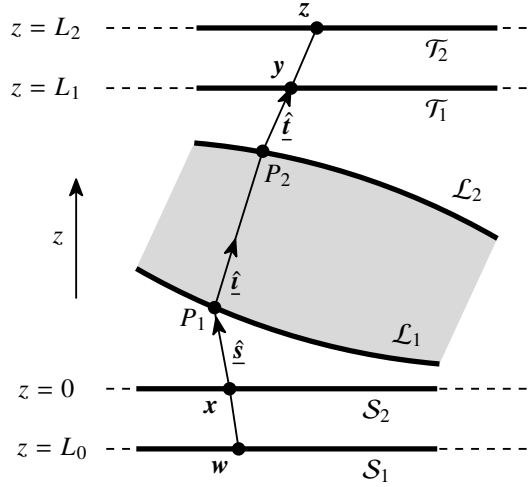
\begin{figure}[!ht]
\begin{center}
\begin{tikzpicture}[scale=0.8]
    \path[fill=gray!30] 
        (0,2) arc (240:265:12.2cm) -- (6,2.5) arc (60:85:12.2cm) -- cycle;
    \draw [line width=0.7mm] (0,-1) -- (5,-1); 
    \draw [line width=0.7mm] (0,0) -- (5,0); 
    \draw [line width=0.7mm] (1,5) -- (6,5); 
    \draw [line width=0.7mm] (1,6) -- (6,6); 
    \draw [line width=0.7mm] (0,2) arc (240:265:12.2cm); 
    \draw [line width=0.7mm] (6,2.5) arc (60:85:12.2cm); 
    \draw [line width=0.25mm] (1.7,-1) -- (1.55,0); 
    \draw [line width=0.25mm] (1.33,1.1) -- (1.3,1.35); 
    \draw [line width=0.25mm] (1.617,2.4) -- (2.1,3.95); 
    \draw [line width=0.25mm] (2.57,5) -- (3,6); 
    \draw [line width=0.25mm, arrows = {-Stealth[length=8pt, inset=2pt,round]}] (1.55,0) -- (1.325,1.2); 
    \draw [line width=0.25mm, arrows = {-Stealth[length=8pt, inset=2pt,round]}] (1.3,1.35) -- (1.65,2.5); 
    \draw [line width=0.25mm, arrows = {-Stealth[length=8pt, inset=2pt,round]}] (2.1,3.95) -- (2.57,5); 
    \draw [line width=0.25mm, arrows = {-Stealth[length=8pt, inset=2pt,round]}] (-1,1.5) -- (-1,3.5); 
    \draw [line width=0.25mm, dashed] (-0.5,-1) -- (6.5,-1);
    \draw [line width=0.25mm, dashed] (-0.5,0) -- (6.5,0);
    \draw [line width=0.25mm, dashed] (-0.5,5) -- (6.5,5);
    \draw [line width=0.25mm, dashed] (-0.5,6) -- (6.5,6);
    \node at (1.7,-1)[circle,fill,inner sep=1.8pt]{}; 
    \node at (1.55,0)[circle,fill,inner sep=1.8pt]{}; 
    \node at (1.3,1.35)[circle,fill,inner sep=1.8pt]{}; 
    \node at (2.1,3.95)[circle,fill,inner sep=1.8pt]{}; 
    \node at (2.57,5)[circle,fill,inner sep=1.8pt]{}; 
    \node at (3,6)[circle,fill,inner sep=1.8pt]{}; 
    \node at (4.5,0.9) {$\mathcal{L}_1$};
    \node at (5.5,3.3) {$\mathcal{L}_2$};
    \node at (4,-1.36) {$\mathcal{S}_1$};
    \node at (4,-0.36) {$\mathcal{S}_2$};
    \node at (5,4.64) {$\mathcal{T}_1$};
    \node at (5,5.64) {$\mathcal{T}_2$};
    \node at (-1.4,-1) {$z=L_0$};
    \node at (-1.5,0) {$z=0$};
    \node at (-1.5,5) {$z=L_1$};
    \node at (-1.5,6) {$z=L_2$};
    \node at (-1.3,2.5) {$z$};
    \node at (1.4,-1.3) {$\bm{w}$};
    \node at (1.25,-0.3) {$\bm{x}$};
    \node at (2.37,5.3) {$\bm{y}$};
    \node at (2.8,6.3) {$\bm{z}$};
    \node at (0.95,1.15) {$P_1$};
    \node at (2.35,3.55) {$P_2$};
    \node at (1.7,0.6) {$\underline{\hat{\bm{s}}}$};
    \node at (1.65,1.75) {$\underline{\hat{\bm{\imath}}}$};
    \node at (2.5,4.35) {$\underline{\hat{\bm{t}}}$};
\end{tikzpicture}
\end{center}
\caption{Sketch of the lens system.}
\label{fig:2S2T_lens_system}
\end{figure}

We consider the lens system illustrated in Fig.~\ref{fig:2S2T_lens_system}, where, similarly as in the reflector system, a light ray propagates from $(\bm{w},L_0)$ on the first source plane $\mathcal{S}_1$ to $(\bm{x},0)$ on the second source plane $\mathcal{S}_2$ in the direction $\underline{\hat{\bm{s}}}$. It reaches the first lens surface $\mathcal{L}_1$ in the point $P_1$ and refracts in direction $\underline{\hat{\bm{\imath}}}$. Next, it reaches the second lens surface $\mathcal{L}_2$ in $P_2$ and refracts in direction $\underline{\hat{\bm{t}}}$. The ray then reaches $(\bm{y},L_1)$ on the first target plane $\mathcal{T}_1$, maintains its direction $\underline{\hat{\bm{t}}}$ and finally reaches $(\bm{z},L_2)$ on the second target plane $\mathcal{T}_2$. Using a similar notation for the reflector system, $u_1=u_1(\bm{x})$ is the distance from $(\bm{x},0)$ to $P_1$, $d=d(\bm{x},\bm{y})$ is the distance from $P_1$ to $P_2$ and $u_2=u_2(\bm{y})$ is the distance from $P_2$ to $(\bm{y},L_1)$. The lens surfaces can be parameterized as
\begin{align}\label{eq:parameterizations_lens}
    \mathcal{L}_1: \underline{\bm{r}}_1=\underline{\bm{r}}_1(\bm{x})=\underline{\bm{x}}+u_1(\bm{x})\underline{\hat{\bm{s}}}\in\mathbb{R}^3,
    &&
    \mathcal{L}_2: \underline{\bm{r}}_2=\underline{\bm{r}}_2(\bm{y})=\underline{\bm{y}}-u_2(\bm{y})\underline{\hat{\bm{t}}}\in\mathbb{R}^3.
\end{align}

\noindent The optical path length $V(\bm{x},\bm{y})$ from $\mathcal{S}_2$ to $\mathcal{T}_1$ is in this case given by
\begin{align}\label{eq:2S2T_OPL_V_lens}
    V=V(\bm{x},\bm{y})=u_1(\bm{x})+nd(\bm{x},\bm{y})+u_2(\bm{y}), &&
    d=d(\bm{x},\bm{y})=\big|\;\underline{\bm{y}}-\underline{\bm{x}}-u_1\underline{\hat{\bm{s}}}-u_2\underline{\hat{\bm{t}}}\;\big|,
\end{align}
with $n>1$ the refractive index of the lens and where we assumed $n_\text{air}=1$ \cite[p. 118]{Hecht2012}. By rewriting the first equation to $d=\frac{1}{n}(V-u_1-u_2)$ and squaring both sides, we find
\begin{align}\label{eq:distance_d_lens_squared}
    d^2=n^{-2}(V^2+u_1^2+u_2^2+2u_1 u_2-2V(u_1+u_2)).
\end{align}
By squaring both sides of the expression for $d$ in Eq.~(\ref{eq:2S2T_OPL_V_lens}) and setting it equal to Eq.~(\ref{eq:distance_d_lens_squared}), we obtain
\begin{align}
    0=&|\underline{\bm{y}}-\underline{\bm{x}}|^2-n^{-2}V^2-2u_1(\underline{\bm{y}}-\underline{\bm{x}})\bm{\cdot}\underline{\hat{\bm{s}}}-2u_2(\underline{\bm{y}}-\underline{\bm{x}})\bm{\cdot}\underline{\hat{\bm{t}}}\nonumber\\
    &+2u_1u_2(\underline{\hat{\bm{s}}}\bm{\cdot}\underline{\hat{\bm{t}}}-n^{-2})+u_1^2(1-n^{-2})+u_2^2(1-n^{-2})+2n^{-2}V(u_1+u_2).\label{eq:V1_squared_big_equation_for_lens}
\end{align}
Notice that when $n=1$, Eq.~(\ref{eq:V1_squared_big_equation_for_lens}) simplifies to Eq.~(\ref{eq:V1_squared}). After multiplying Eq.~(\ref{eq:V1_squared_big_equation_for_lens}) by $n^2$, we can write it as
\begin{align}\label{eq:quadratic_eq_for_u1}
    b_0 u_1^2+ 2b_1 u_1 + b_2 = 0,
\end{align}
or as
\begin{align}\label{eq:quadratic_eq_for_u2}
    b_0 u_2^2+ 2b_3 u_2 + b_4 = 0,
\end{align}
with auxiliary variables
\begin{subequations}
\label{eq:auxiliary_variables_lens}
\begin{align}
    b_0&=n^2-1,\\
    b_1&=V-n^2(\underline{\bm{y}}-\underline{\bm{x}})\bm{\cdot}\underline{\hat{\bm{s}}}+u_2(n^2\underline{\hat{\bm{s}}}\bm{\cdot}\underline{\hat{\bm{t}}}-1),\\
    b_2&=n^2|\underline{\bm{y}}-\underline{\bm{x}}|^2-V^2-2n^2 u_2(\underline{\bm{y}}-\underline{\bm{x}})\bm{\cdot}\underline{\hat{\bm{t}}}+(n^2-1)u_2^2+2Vu_2,\\
    b_3&=V-n^2(\underline{\bm{y}}-\underline{\bm{x}})\bm{\cdot}\underline{\hat{\bm{t}}}+u_1(n^2\underline{\hat{\bm{s}}}\bm{\cdot}\underline{\hat{\bm{t}}}-1),\\
    b_4&=n^2|\underline{\bm{y}}-\underline{\bm{x}}|^2-V^2-2n^2u_1(\underline{\bm{y}}-\underline{\bm{x}})\bm{\cdot}\underline{\hat{\bm{s}}}+(n^2-1)u_1^2+2Vu_1.
\end{align}
\end{subequations}
From the quadratic equation (\ref{eq:quadratic_eq_for_u1}) we compute $u_1(\bm{x})=G_\text{L}(\bm{x},\bm{y},u_2(\bm{y});V(\bm{x},\bm{y}))$ with generating function $G_\text{L}(\bm{x},\bm{y},w;V)$ defined as
\begin{subequations}
\label{eq:generating_functions_lens}
\begin{align}\label{eq:generating_functions_lens_G}
    G_\text{L}=G_\text{L}(\bm{x},\bm{y},u_2(\bm{y});V(\bm{x},\bm{y}))=-\frac{1}{b_0}\Big(b_1+\sqrt{b_1^2-b_0b_2}\Big).
\end{align}
In addition, from quadratic equation (\ref{eq:quadratic_eq_for_u2}) we compute $u_2(\bm{y})=H_\text{L}(\bm{x},\bm{y},u_1(\bm{x});V(\bm{x},\bm{y}))$ with inverse generating function $H_\text{L}(\bm{x},\bm{y},w;V)$ defined as
\begin{align}\label{eq:generating_functions_lens_H}
    H_\text{L}=H_\text{L}(\bm{x},\bm{y},u_1(\bm{x});V(\bm{x},\bm{y}))=-\frac{1}{b_0}\Big(b_3+\sqrt{b_3^2-b_0b_4}\Big).
\end{align}
\end{subequations}
By substituting Eqs. (\ref{eq:auxiliary_variables_lens}) and $V$ from Eqs. (\ref{eq:2S2T_OPL_V_lens}), we can show that the arguments of the square roots in expressions (\ref{eq:generating_functions_lens_G}) and (\ref{eq:generating_functions_lens_H}) are strictly positive. This means that there is no internal reflection, when choosing the minus-sign solution of the quadratic equations (\ref{eq:quadratic_eq_for_u1}) and (\ref{eq:quadratic_eq_for_u2}).

\subsection{Energy conservation}
We will now derive relations for energy conservation in the form of \textit{Jacobian equations}: one between $\mathcal{S}_1$ and $\mathcal{S}_2$, another between $\mathcal{S}_2$ and $\mathcal{T}_1$, and a third between $\mathcal{T}_1$ and $\mathcal{T}_2$. To this end, we define optical mappings $\bm{m}_\mathcal{S}:\mathcal{S}_2\to\mathcal{S}_1$, $\bm{m} :\mathcal{S}_2\to\mathcal{T}_1$ and $\bm{m}_\mathcal{T}:\mathcal{T}_1\to\mathcal{T}_2$ by $\bm{m}_\mathcal{S}(\bm{x})=\bm{w}$, $\bm{m} (\bm{x})=\bm{y}$ and $\bm{m}_\mathcal{T}(\bm{y})=\bm{z}$ with compact domains $\mathcal{S}_1$, $\mathcal{S}_2$, $\mathcal{T}_1$ and $\mathcal{T}_2$. While mapping $\bm{m}_\mathcal{S}$ could alternatively be defined with domain $\mathcal{S}_1$ and range $\mathcal{S}_2$, we adopt the current definition to simplify notation throughout the paper. We assume that $\mathcal{S}_1$ emits light with emittance $f_1:\mathcal{S}_1\to(0,\infty)$, which is then sent to $\mathcal{S}_2$ where the light is distributed according to $f_2:\mathcal{S}_2\to(0,\infty)$. The light then reaches $\mathcal{T}_1$ with illuminance $g_1:\mathcal{T}_1\to(0,\infty)$ and finally $\mathcal{T}_2$ with illuminance $g_2:\mathcal{T}_2\to(0,\infty)$. From energy conservation between $\mathcal{S}_2$ and $\mathcal{T}_1$ we have
\begin{align*}
    \iint_{\mathcal{A}}f_2(\bm{x})\;\mathrm{d}\bm{x}=\iint_{\bm{m} (\mathcal{A})}g_1(\bm{y})\;\mathrm{d}\bm{y}=\iint_{\mathcal{A}}g_1(\bm{m} (\bm{x}))\;|\text{det}(\text{D}\bm{m} (\bm{x}))|\;\mathrm{d}\bm{x},
\end{align*}
for arbitrary $\mathcal{A}\subset \mathcal{S}_2$, where $\text{D}\bm{m} (\bm{x})$ denotes the Jacobian matrix of $\bm{m} (\bm{x})$. By assuming $\text{det}(\text{D}\bm{m} (\bm{x}))>0$, we obtain the Jacobian equation
\begin{subequations}
\label{eq:2S2T_Jacobian_equations}
\begin{align}\label{eq:2S2T_Jacobian_equations_1}
    \det(\text{D}\bm{m} (\bm{x}))&=F(\bm{x},\bm{m} (\bm{x})),& F(\bm{x},\bm{m} (\bm{x}))&:=\frac{f_2(\bm{x})}{g_1(\bm{m} (\bm{x}))},
\end{align}
which we will use for computing the mapping $\bm{y}=\bm{m} (\bm{x})$. Similarly, by imposing energy conservation between $\mathcal{S}_2$ and $\mathcal{S}_1$ with the assumption $\text{det}(\text{D}\bm{m}_\mathcal{S}(\bm{x}))>0$, and energy conservation between $\mathcal{T}_1$ and $\mathcal{T}_2$ with the assumption $\text{det}(\text{D}\bm{m}_\mathcal{T}(\bm{y}))>0$, we find the Jacobian equations
\begin{align}
    \text{det}(\text{D}\bm{m}_\mathcal{S}(\bm{x}))=F_\mathcal{S}(\bm{x},\bm{m}_\mathcal{S}(\bm{x})), 
    &&
    F_\mathcal{S}(\bm{x},\bm{m}_\mathcal{S}(\bm{x}))&:=\frac{f_2(\bm{x})}{f_1(\bm{m}_\mathcal{S}(\bm{x}))},\\
    \text{det}(\text{D}\bm{m}_\mathcal{T}(\bm{y}))=F_\mathcal{T}(\bm{y},\bm{m}_\mathcal{T}(\bm{y})),
    &&
    F_\mathcal{T}(\bm{y},\bm{m}_\mathcal{T}(\bm{y}))&:=\frac{g_1(\bm{y})}{g_2(\bm{m}_\mathcal{T}(\bm{y}))},
\end{align}
\end{subequations}
which we will use for computing $\bm{w}=\bm{m}_\mathcal{S}(\bm{x})$ and $\bm{z}=\bm{m}_\mathcal{T}(\bm{y})$ \cite{Villani2003}.

In addition to the energy balances, we impose boundary conditions. Since all light originating from $\mathcal{S}_1$ should reach $\mathcal{S}_2$, $\mathcal{T}_1$ and $\mathcal{T}_2$, we require $\bm{m}_\mathcal{S}(\mathcal{S}_2)=\mathcal{S}_1$, $\bm{m} (\mathcal{S}_2)=\mathcal{T}_1$ and $\bm{m}_\mathcal{T}(\mathcal{T}_1)=\mathcal{T}_2$. To transform these into boundary conditions, we use the edge-ray principle \cite{Ries1994,Winston2005}, which states that these conditions are equivalent to the transport boundary conditions
\begin{subequations}
\begin{align}
    \bm{m}_\mathcal{S}(\partial \mathcal{S}_2)&=\partial \mathcal{S}_1,
    \label{eq:transport_boundary_condition_BC_0}
    \\
    \bm{m} (\partial \mathcal{S}_2)&=\partial \mathcal{T}_1,\label{eq:transport_boundary_condition_BC_1}\\
    \bm{m}_\mathcal{T}(\partial \mathcal{T}_1)&=\partial \mathcal{T}_2,\label{eq:transport_boundary_condition_BC_2}
\end{align}
\end{subequations}
see \cite[p. 65-66]{Teun2024}.

\subsection{Optical path length V}\label{sec:Hamilton2}
Under the assumption that the optical mappings $\bm{w}=\bm{m}_\mathcal{S}(\bm{x})$ and $\bm{z}=\bm{m}_\mathcal{T}(\bm{y})$ are known, a light ray's optical momentum at $\mathcal{S}_2$ and $\mathcal{T}_1$ is given by
\begin{align}\label{eq:momentum}
    \bm{p}_\mathrm{s}(\bm{x})
    =
    \frac{\bm{x}-\bm{m}_\mathcal{S}(\bm{x})}{\sqrt{|\bm{x}-\bm{m}_\mathcal{S}(\bm{x})|^2+L_0^2}},&&
    \bm{p}_\mathrm{t}(\bm{y})
    =
    \frac{\bm{m}_\mathcal{T}(\bm{y})-\bm{y}}{\sqrt{|\bm{m}_\mathcal{T}(\bm{y})-\bm{y}|^2+(L_2-L_1)^2}}.
\end{align}
We can calculate the optical path length $V$ from $\mathcal{S}_2$ to $\mathcal{T}_1$ by using these expressions in combination with the equations
\begin{align}\label{eq:OPL_V_properties}
    \nabla_{\bm{x}}V=-\bm{p}_\mathrm{s}(\bm{x}), &&
    \nabla_{\bm{y}}V=\bm{p}_\mathrm{t}(\bm{y}),
\end{align}
which can be derived using Hamilton's characteristic functions \cite[p. 142]{Born1999}. We are only interested in $V(\bm{x},\bm{y})$ for $\bm{y}=\bm{m} (\bm{x})$. Therefore, we define $\widetilde{V}=\widetilde{V}(\bm{x})=V(\bm{x},\bm{m} (\bm{x}))=V(x_1,x_2,m_1(x_1,x_2),m_2(x_1,x_2))$ with gradient
\begin{align}\nonumber
    \nabla_{\bm{x}}\widetilde{V}&=\begin{pmatrix}
        \frac{\partial V}{\partial x_1}+\frac{\partial V}{\partial y_{1}}\frac{\partial m_{1}}{\partial x_1}+\frac{\partial V}{\partial y_{2}}\frac{\partial m_{2}}{\partial x_1}
        \vspace{2pt}
        \\
        \frac{\partial V}{\partial x_2}+\frac{\partial V}{\partial y_{1}}\frac{\partial m_{1}}{\partial x_2}+\frac{\partial V}{\partial y_{2}}\frac{\partial m_{2}}{\partial x_2}
    \end{pmatrix}
    =
    \begin{pmatrix}
        \frac{\partial V}{\partial x_1}
        \\
        \frac{\partial V}{\partial x_2}
    \end{pmatrix}
    +
    \begin{pmatrix}
        \frac{\partial m_{1}}{\partial x_1} & \frac{\partial m_{2}}{\partial x_1}\\
        \frac{\partial m_{1}}{\partial x_2} & \frac{\partial m_{2}}{\partial x_2}
    \end{pmatrix}
    \begin{pmatrix}
        \frac{\partial V}{\partial y_{1}}\\
        \frac{\partial V}{\partial y_{2}}
    \end{pmatrix}
    \\
    &=
    \nabla_{\bm{x}}V+(\text{D}\bm{m} )^\text{T}\nabla_{\bm{y}}V=-\bm{p}_\mathrm{s}(\bm{x})+(\text{D}\bm{m} )^\text{T}\bm{p}_\mathrm{t}(\bm{m} (\bm{x})).\label{eq:gradient_V_tilde}
\end{align}
We will use this equation to determine $\widetilde{V}$.

\subsection{$H$-convex analysis}\label{sec:convexity_crit}
Generating functions $G$ and $H$ for the reflector system ($G_\text{R}$ and $H_\text{R}$) and lens system ($G_\text{L}$ and $H_\text{L}$) are defined such that $u_1(\bm{x})=G(\bm{x},\bm{y},u_2(\bm{y});V(\bm{x},\bm{y}))$ and $u_2(\bm{y})=H(\bm{x},\bm{y},u_1(\bm{x}),V(\bm{x},\bm{y}))$. There are many pairs $(u_1(\bm{x}),u_2(\bm{y}))$ satisfying these equations. Since we can verify that $H_w<0$ for Eqs.~(\ref{eq:2T_generating_function_G}) and (\ref{eq:generating_functions_lens_G}), a possible solution is given by the $G-$convex $H-$convex pair
\begin{subequations}
\begin{align}
    \forall \bm{x}\in\mathcal{S}_2\;\;\; u_1(\bm{x})&=\max_{\bm{y}\in\mathcal{T}_1}G(\bm{x},\bm{y},u_2(\bm{y}),V(\bm{x},\bm{y})),\label{eq:maximization_G}\\
    \forall \bm{y}\in\mathcal{T}_1\;\;\; u_2(\bm{y})&=\max_{\bm{x}\in\mathcal{S}_2}H(\bm{x},\bm{y},u_1(\bm{x}),V(\bm{x},\bm{y})).\label{eq:maximization_H}
\end{align}
\end{subequations}
The value $\bm{y}\in\mathcal{T}_1$ that maximizes $G$ in Eq.~(\ref{eq:maximization_G}) defines $\bm{y}=\bm{m} (\bm{x})$, implying that $\widetilde{H}(\bm{x},\bm{y})=H(\bm{x},\bm{y},u_1(\bm{x});V(\bm{x},\bm{y}))$ has a stationary point at $\bm{y}=\bm{m} (\bm{x})$. Thus,
\begin{align}
    \nabla_{\bm{x}}\widetilde{H}(\bm{x},\bm{y})=\;&\nabla_{\bm{x}}H(\bm{x},\bm{y},u_1(\bm{x}),V(\bm{x},\bm{y}))\nonumber\\
    &+H_w(\bm{x},\bm{y},u_1(\bm{x}),V(\bm{x},\bm{y}))\nabla u_1(\bm{x})\nonumber\\
    &+H_V(\bm{x},\bm{y},u_1(\bm{x}),V(\bm{x},\bm{y}))\nabla_{\bm{x}} V(\bm{x},\bm{y})=\bm{0}.\label{eq:gradient_H_tilde}
\end{align}
Provided $\nabla u_{1}$ and $\nabla_{\bm{x}}V$ are known, the implicit function theorem implies that this equation provides a mapping $\bm{y}=\bm{m} (\bm{x})$ under the condition that the mixed Hessian matrix
\begin{align}\label{eq:matrix_C}
    \bm{C}=\bm{C}(\bm{x},\bm{y},u_1(\bm{x});V(\bm{x},\bm{y}))=\text{D}_{\bm{xy}}\widetilde{H}(\bm{x},\bm{y})=\left(\frac{\partial^2\widetilde{H}(\bm{x},\bm{y})}{\partial x_i\,\partial y_j}\right),
\end{align}
is invertible for all $\bm{x}\in\mathcal{S}_2$ and $\bm{y}\in\mathcal{T}_1$, which we verified for our numerical simulations. After substituting $\bm{y}=\bm{m}(\bm{x})$ in Eq.~(\ref{eq:gradient_H_tilde}) and taking the derivative with respect to $\bm{x}$, we have
\begin{align}\label{eq:not_matrix_C}
    \text{D}_{\bm{xx}}\widetilde{H}(\bm{x},\bm{m} (\bm{x}))+\text{D}_{\bm{xy}}\widetilde{H}(\bm{x},\bm{m} (\bm{x}))\text{D}\bm{m} (\bm{x})=\bm{O},
\end{align}
which we rewrite to
\begin{align}\label{eq:CDmP}
    \bm{C}\text{D}\bm{m} =\bm{P}, && \bm{P}=\bm{P}(\bm{x},\bm{y},u_1(\bm{x});V(\bm{x},\bm{y})):=-\text{D}_{\bm{xx}}\widetilde{H}(\bm{x},\bm{y}),
\end{align}
where $\text{D}_{\bm{xx}}\widetilde{H}$ denotes the Hessian matrix of $\widetilde{H}$ with respect to $\bm{x}$. A sufficient condition for the maximum in Eq.~(\ref{eq:maximization_H}) is for $\bm{P}$ to be symmetric positive definite (SPD), satisfying
\begin{align}\label{eq:detP1_constraint}
    \text{det}(\bm{P})=F(\cdot,\bm{m})\text{det}(\bm{C}),
\end{align}
where $F(\cdot,\bm{m})$ is given in Eq.~(\ref{eq:2S2T_Jacobian_equations_1}) \cite[p. 124]{Lotte2021}.

\subsection{A far-field target with stereographic projection}\label{sec:FF_target}
Although both the reflector and lens systems have two targets, target $\mathcal{T}_2$ can be replaced by a far-field target that specifies the target direction of the light rays. To this end, consider a fixed target vector $\underline{\hat{\bm{t}}}$ which originates from point $(\bm{y},L_1)$ on the first target $\mathcal{T}_1$ and is directed towards the point $(\bm{z},L_2)$ on the second target $\mathcal{T}_2$. A \textit{generalized gnomonic projection} is the relation between $\Delta \bm{y}:=\bm{z}-\bm{y}$ and $\underline{\hat{\bm{t}}}$, and can be written as 
\begin{align}\label{eq:gnom_coord}
    \Delta \bm{y}=\frac{L_2-L_1}{t_3}\bm{p}_\mathrm{t},
\end{align}
which is undefined for $t_3=0$. Alternatively, a \textit{stereographic projection} from the south pole projects target vector $\underline{\hat{\bm{t}}}$, represented by a point on the unit sphere, onto the plane $z=0$, defined by
\begin{align}
    \bm{\mathcal{P}}_\text{ster}[\underline{\hat{\bm{t}}}]=\frac{1}{1+ t_3}\bm{p}_\mathrm{t},
\end{align}
which is undefined for $t_3=-1$. Generalized gnomonic projections~(\ref{eq:gnom_coord}) can be expressed in stereographic projections by the formula
\begin{align}
    \Delta\bm{y}=\frac{2(L_2-L_1)\bm{\mathcal{P}}_\text{ster}[\underline{\hat{\bm{t}}}]}{1-|\bm{\mathcal{P}}_\text{ster}[\underline{\hat{\bm{t}}}]|^2}.
\end{align}
Thus, we can replace target $\mathcal{T}_2$ by a far-field target using stereographic coordinates to find the target direction $\underline{\hat{\bm{t}}}$ of light rays and then convert them back to positional generalized gnomonic coordinates on target $\mathcal{T}_2$.




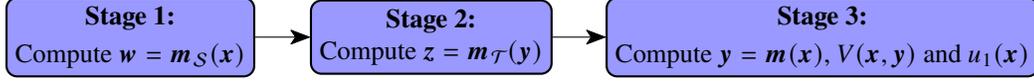
\begin{figure}[!ht]
\begin{center}
    \begin{tikzpicture}[node distance=0.8cm]
        \node (step1) [process] {\textbf{Stage 1:}\\[0.2em]
            Compute $\bm{w}=\bm{m}_\mathcal{S}(\bm{x})$};
        
        \node (step2) [process, right of=step1, yshift=0cm, xshift = 3.2cm] {\textbf{Stage 2:}\\
        Compute $\bm{z}=\bm{m}_\mathcal{T}(\bm{y})$};

        \node (step3) [process, right of=step2, yshift=0cm, xshift = 4.4cm] {\textbf{Stage 3:}\\[0.2em]
        Compute $\bm{y}=\bm{m}(\bm{x})$, $V(\bm{x},\bm{y})$ and $u_1(\bm{x})$};

        \draw [arrows = {-Stealth[length=8pt, inset=2pt,round]}] (step1) -- (step2);
        \draw [arrows = {-Stealth[length=8pt, inset=2pt,round]}] (step2) -- (step3);
    \end{tikzpicture}
\end{center}
\caption{Flowchart of the least-squares algorithm.}
\label{fig:flowchart_LSS}
\end{figure}

\section{The least-squares algorithm}\label{sec:Algorithm}
To compute the optical surfaces in the reflector and lens systems, we apply the three-stage least-squares algorithm illustrated in Fig. \ref{fig:flowchart_LSS}. We will first discuss Stage 3 in detail. Stages 1 and 2 are simpler cases, as we will show at the end of this section.

\textbf{Stage 3:} The final stage of the algorithm iteratively computes the mapping $\bm{y}=\bm{m} (\bm{x})$, the optical path length $V=V(\bm{x},\bm{y})$ and the reflector surfaces. To compute $\bm{y}=\bm{m} (\bm{x})$, it starts from an initial guess $\bm{m} ^0$, which maps the smallest bounding box enclosing $\mathcal{S}_2$ to the smallest bounding box enclosing $\mathcal{T}_1$. Subsequently, the algorithm minimizes the functionals
\begin{subequations}
\begin{align}
    J_{\mathrm{I}}[\bm{m} ,\bm{P}]&=\tfrac{1}{2}\iint_{\mathcal{S}_2}||\bm{C}\text{D}\bm{m} -\bm{P}||_\mathrm{F}^2\;\mathrm{d}\bm{x},\label{eq:functional_J_I}
    \\
    J_\mathrm{B}[\bm{m} ,\bm{b}]&=\tfrac{1}{2}\oint_{\partial \mathcal{S}_2}|\bm{m} -\bm{b}|^2\;\mathrm{d}s,\label{eq:functional_J_B}
    \\
    J[\bm{m} ,\bm{P},\bm{b}]&=\alpha J_\mathrm{I}[\bm{m} ,\bm{P}]+(1-\alpha)J_\mathrm{B}[\bm{m} ,\bm{b}],\label{eq:functional_J}
    \\
    K[\bm{m} ,V]&=\tfrac{1}{2}\iint_{\mathcal{S}_2}|\nabla_{\bm{x}} V(\bm{x},\bm{m})+\bm{p}_\mathrm{s}(\bm{x})-(\mathrm{D}\bm{m})^T\bm{p}_\mathrm{t}(\bm{m})|^2\;\mathrm{d}\bm{x},
    \\
    I[\bm{m} ,u_1,V]&=\tfrac{1}{2}\iint_{\mathcal{S}_2}\Big|\frac{\nabla_{\bm{x}} H(\bm{x},\bm{m} ,u_1,V)+H_V(\bm{x},\bm{m} ,u_1,V)\nabla_{\bm{x}}V(\bm{x},\bm{m})}{H_w(\bm{x},\bm{m} ,u_1,V)}+\nabla u_1\Big|^2 \mathrm{d}\bm{x},\label{eq:functional_I}
\end{align}
where $||\cdot||_\mathrm{F}$ denotes the Frobenius norm, by computing
\begin{align}
    \bm{P}^{k+1}&=\underset{\bm{P}\in\mathcal{P}}{\mathrm{argmin}}\;J_\mathrm{I}[\bm{m} ^k,\bm{P}],\label{eq:computing_1}
    \\
    \bm{b}^{k+1}&=\underset{\bm{b}\in\mathcal{B}}{\mathrm{argmin}}\;J_\mathrm{B}[\bm{m} ^k,\bm{b}],\label{eq:computing_2}
    \\
    \bm{m} ^{k+1}&=\underset{\bm{m} \in C^2(\mathcal{S}_2)^2}{\mathrm{argmin}}\;J[\bm{m} ,\bm{P}^{k+1},\bm{b}^{k+1}],\label{eq:computing_3}
    \\
    V^{k+1}&=\underset{V\in C^2(\mathcal{S}_2)}{\mathrm{argmin}}\;K[\bm{m} ^{k+1},V],\label{eq:computing_4}
    \\
    u_1^{k+1}&=\underset{u_1\in C^2(\mathcal{S}_2)}{\mathrm{argmin}}\;I[\bm{m} ^{k+1},u_1,V^{k+1}],\label{eq:computing_5}
\end{align}
with function spaces
\begin{align}
    \mathcal{P}&=\{\bm{P}\in C^1(\mathcal{S}_2)^{2\times2}\mid \bm{P}\text{ is SPD, }\text{det}(\bm{P})=F(\cdot,\bm{m}\cdot)\text{det}(\bm{C})\},
    \\
    \mathcal{B}&=\{\bm{b}\in C^1(\partial \mathcal{S}_2)^2\mid\bm{b}(\bm{x})\in\partial\mathcal{T}_1\}.
\end{align}
\label{eq:iteration_scheme}\end{subequations}
\noindent In the first two steps of this scheme, the functionals $J_\mathrm{I}$ and $J_\mathrm{B}$ are minimized for $\bm{P}$ and $\bm{b}$ to satisfy Eq.~(\ref{eq:CDmP}) under the constraint (\ref{eq:detP1_constraint}) and the transport boundary condition (\ref{eq:transport_boundary_condition_BC_1}). The third step of the scheme then minimizes a linear combination $J$ with weighing factor $\alpha\in(0,1)$ of both functionals $J_\mathrm{I}$ and $J_\mathrm{B}$ to update the mapping $\bm{y}=\bm{m} (\bm{x})$. The fourth step of the scheme minimizes the functional $K$ to update $V$ and satisfy Eq.~(\ref{eq:gradient_V_tilde}). Finally, the fifth step minimizes the functional $I$ to update $u_1$ and satisfy Eq.~(\ref{eq:gradient_H_tilde}). At the end of each iteration, we update $u_2(\bm{y})=H(\bm{x},\bm{y},u_1(\bm{x});V(\bm{x},\bm{y}))$ and matrices $\bm{C}$ and $\bm{P}$ using Eqs.~(\ref{eq:matrix_C}) and (\ref{eq:CDmP}). These steps are then repeated for $k=0,1,2,...$. Finally, we use Eqs. (\ref{eq:parameterizations_reflectors}) or (\ref{eq:parameterizations_lens}) to calculate the shapes of the reflectors or lenses.

An in-depth analysis for the minimization of the functionals $J_\mathrm{I}$, $J_\mathrm{B}$ and $J$ using finite volume methods can be found in Romijn \textit{et al.}\ \cite{Lotte2021}. A discussion on the minimization of functionals $K$ and $I$ can be found in Braam \textit{et al.}\ \cite{Braam2025}, which converts the minimization problem to a boundary value problem. We enforce unique solutions by setting the values $V$ and $u_1$ of the center rays equal to $V_0,u_{1,0}\in\mathbb{R}$, respectively.


Notably, in \cite{Braam2025}, we considered a specific case of the reflector system where the source was assumed to be parallel. The first of Eqs.~(\ref{eq:OPL_V_properties}) then implies that the optical path length is independent of the source coordinates, so that $V=V(\bm{y})$. As a result, its calculation can be done in a separate stage in the least-squares algorithm. For the general sources discussed in this article, the optical path length is a function of both $\bm{x}$ and $\bm{y}$ and its calculation can therefore not be decoupled from the final stage of the algorithm.

\textbf{Stages 1 and 2:} The computation of $\bm{w}=\bm{m}_\mathcal{S}(\bm{x})$ and $\bm{z}=\bm{m}_\mathcal{T}(\bm{y})$ resembles the procedure discussed in Stage 3. In this case, however, there are no optical surfaces between the source planes and target planes and therefore, we choose the generating functions
\begin{align}
    G_\mathcal{S}(\bm{x},\bm{y},w)=\bm{x}\bm{\cdot}\bm{y}+w, && G_\mathcal{T}(\bm{x},\bm{y},w)=\bm{x}\bm{\cdot}\bm{y}+w,
\end{align}
with inverse generating functions
\begin{align}\label{eq:stage12revisited_H}
    H_\mathcal{S}(\bm{x},\bm{y},w)=-\bm{x}\bm{\cdot}\bm{y}+w, && H_\mathcal{T}(\bm{x},\bm{y},w)=-\bm{x}\bm{\cdot}\bm{y}+w,
\end{align}
corresponding to a quadratic cost function, for which a derivation can be found in \cite[p. 71]{Lotte2021}. To compute both $\bm{w}=\bm{m}_\mathcal{S}(\bm{x})$ and $\bm{z}=\bm{m}_\mathcal{T}(\bm{y})$, we again use the iteration scheme (\ref{eq:iteration_scheme}) subject to the following modifications:
\begin{itemize}
    \item In each of the expressions in the iteration scheme, $\bm{m}$ is replaced by $\bm{m}_\mathcal{S}$ or $\bm{m}_\mathcal{T}$, $\mathcal{S}_2$ is replaced by $\mathcal{S}_1$ or $\mathcal{T}_1$, $\mathcal{T}_1$ is replaced by $\mathcal{S}_2$ or $\mathcal{T}_2$, $H(\bm{x},\bm{y},w;V)$ is replaced by $H_\mathcal{S}(\bm{x},\bm{y},w)$ or $H_\mathcal{T}(\bm{x},\bm{y},w)$, and $F$ is replaced by $F_\mathcal{S}$ or $F_\mathcal{T}$.
    \item The function $\widetilde{H}(\bm{x},\bm{y})$, used in expressions (\ref{eq:matrix_C}) and (\ref{eq:not_matrix_C}) for matrices $\bm{C}$ and $\bm{P}$, is defined in terms of inverse generating functions $H_\mathcal{S}$ or $H_\mathcal{T}$.
\end{itemize}

\section{Numerical results}\label{sec:results}
In this section we provide two examples for the model from Sec.~\ref{sec:Formulation} which we solve using the least-squares algorithm discussed in Sec.~\ref{sec:Algorithm}. The first example discusses the reflector system and shows that the system can handle point sources and far-field targets. The second example discusses the lens system and shows that the algorithm can create two distinct complex light distributions at the targets.

\begin{figure}[!ht]
\centering
\begin{subfigure}[b]{0.3\textwidth}
    \centering
    \includegraphics[width=\textwidth,trim={0cm 0cm 0cm 0cm},clip]{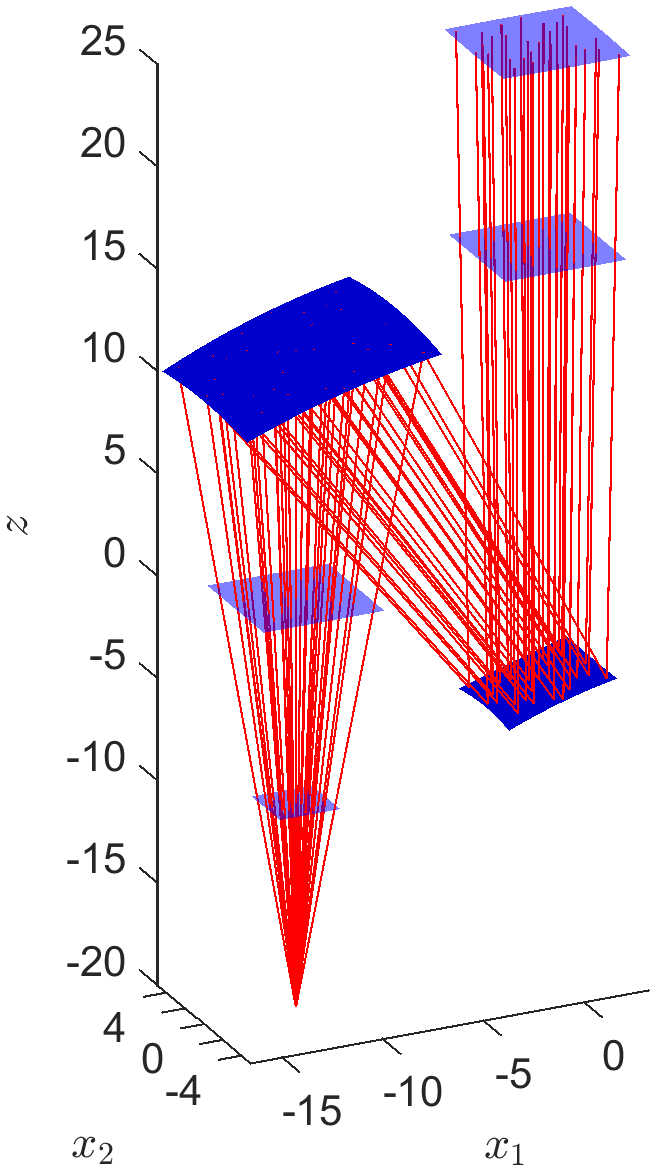}
    \caption{Reflector system.}
    \label{fig:ReflectorSystem}
\end{subfigure}
\hfill
\begin{subfigure}[b]{0.65\textwidth}
    \centering
    \begin{subfigure}[b]{0.48\textwidth}
        \centering
        \includegraphics[width=\textwidth,trim={0cm 0cm 0cm 0cm},clip]{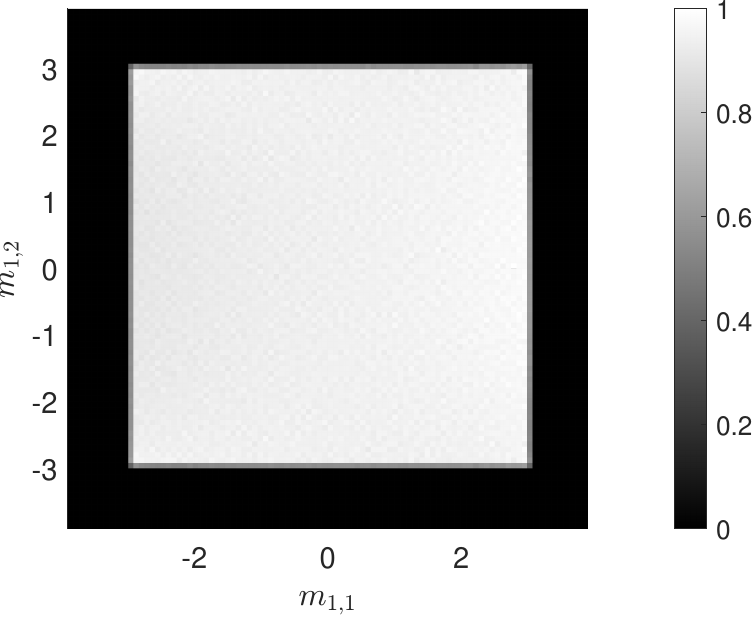}
        \caption{Ray-traced illuminance pattern on $\mathcal{T}_1$.}
        \label{fig:T1_illuminance}
    \end{subfigure}
    \hfill
    \begin{subfigure}[b]{0.503\textwidth}
        \centering
        \includegraphics[width=\textwidth,trim={0cm 0cm 0cm 0cm},clip]{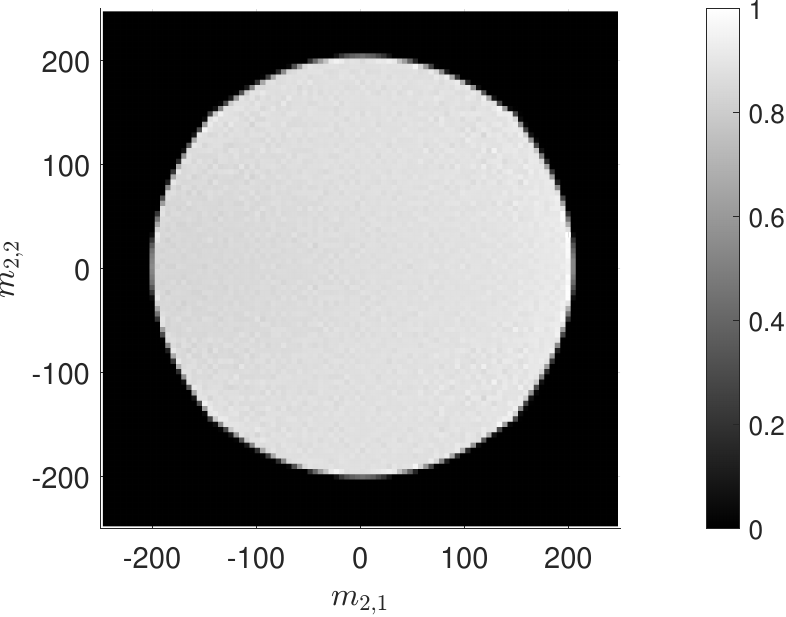}
        \caption{Ray-traced illuminance pattern in the far-field at $z=10^4$.}
        \label{fig:FF_illuminance}
    \end{subfigure}
    \begin{subfigure}[b]{0.48\textwidth}
        \centering
        \includegraphics[width=\textwidth,trim={0cm 0cm 0cm 0cm},clip]{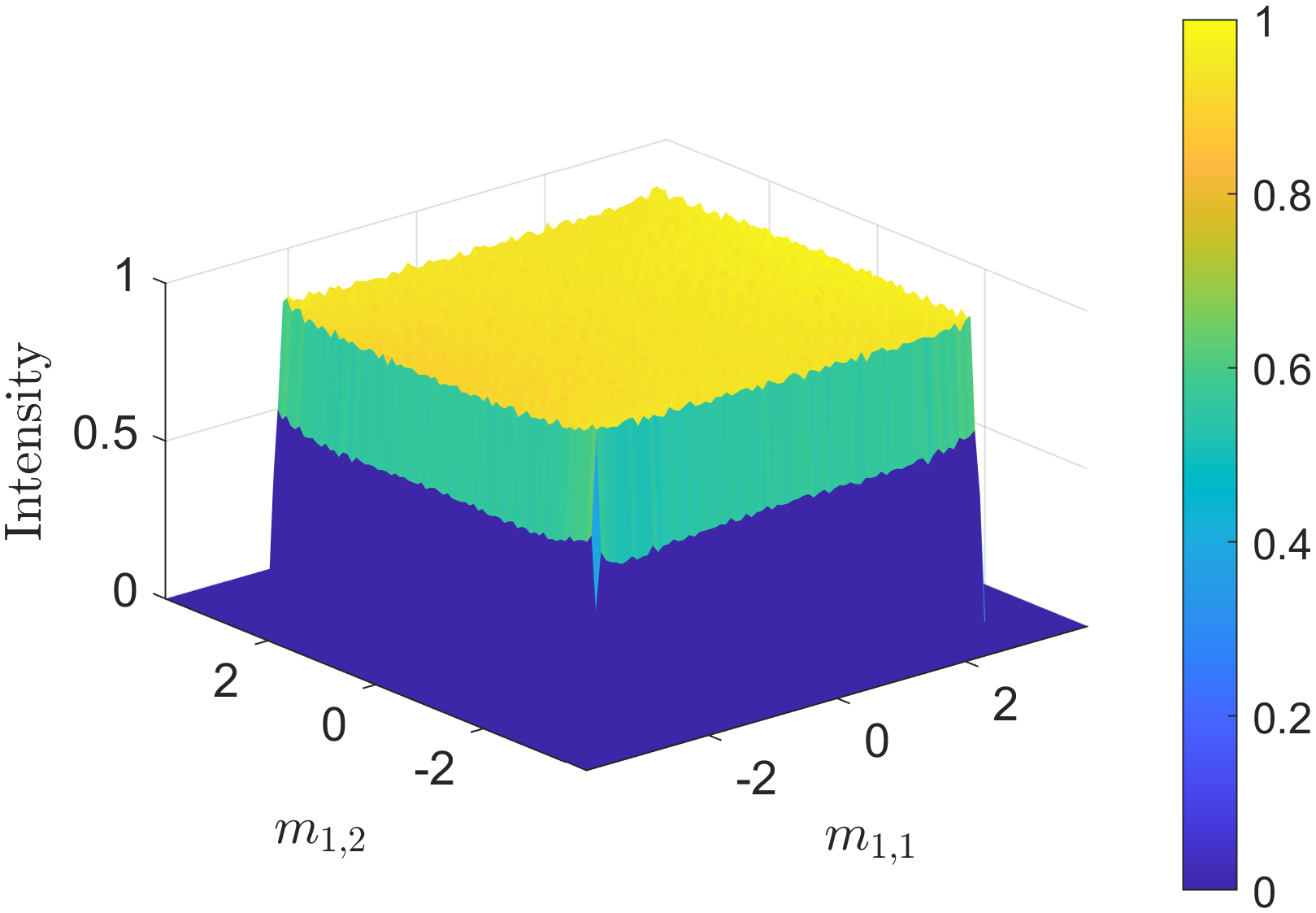}
        \caption{Three-dimensional illuminance pattern on $\mathcal{T}_1$.}
        \label{fig:T1_intensity}
    \end{subfigure}
    \hfill
    \begin{subfigure}[b]{0.48\textwidth}
        \centering
        \includegraphics[width=\textwidth,trim={0cm 0cm 0cm 0cm},clip]{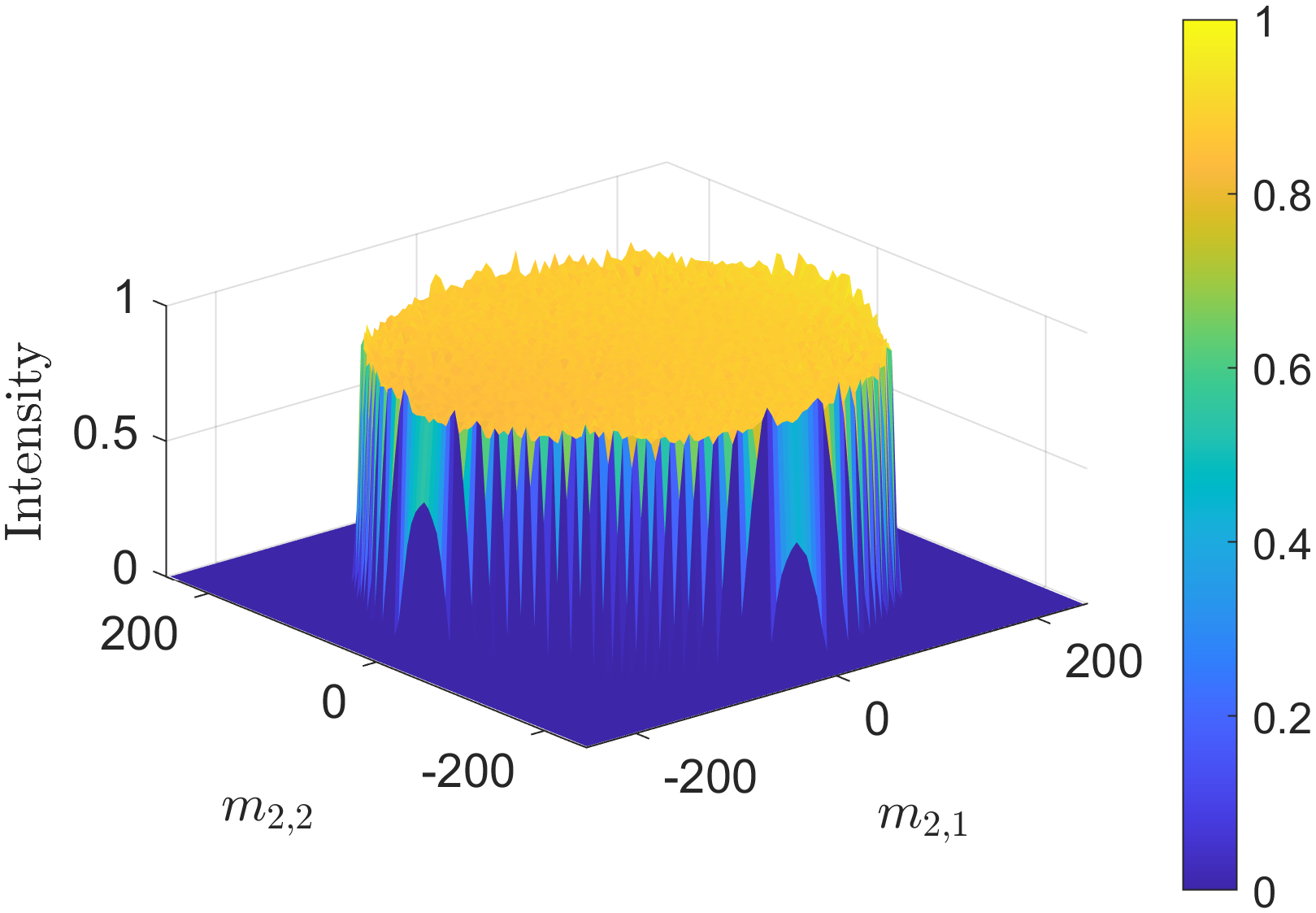}
        \caption{Three-dimensional illuminance pattern in the far-field at $z=10^4$.}
        \label{fig:FF_intensity}
    \end{subfigure}
\end{subfigure}
\caption{Results of the reflector system with $L_0=-10$, $L_1=15$, $L_2=25$, $V_0=55$, $u_{1,0}=12$, computed after $50$ iterations on a $1000\times1000$ grid and verified by ray-tracing using $10^6$ rays and $100\times100$ bins.}
\label{fig:ReflectorResults}
\end{figure}

\subsection{Reflector system}
We consider the reflector system with Cartesian coordinates $\bm{w}\in[-13.5,-10.5]\times[-1.5,1.5]$, $\bm{x}\in[-15,-9]\times[-3,3]$ and $\bm{y}\in[-3,3]\times[-3,3]$, and with stereographic coordinates \mbox{$\bm{z}_{\text{ster}}\in\{\bm{z}\in\mathbb{R}\mid |\bm{z}|\leq 10^{-2}\}$} for a far-field target. We set $L_0=-10$, $L_1=15$, $L_2=25$, $V_0=55$, $u_{1,0}=12$ and take uniformly discretized grids of size $1000\times1000$. We assume that the light at $\mathcal{S}_1$, $\mathcal{S}_2$ and $\mathcal{T}_1$ is uniformly distributed in Cartesian coordinates and that the light in the far-field is uniformly distributed in stereographic coordinates.

After $50$ iterations over each of the three stages in the algorithm with $\alpha=0.5$ during the first two stages and $\alpha=0.01$ during the last stage, it finds the reflector system in Fig.~\ref{fig:ReflectorSystem}. The computation time, on a laptop with Intel Core i7-11800 H 2.30 GHz processor and a RAM of 16.0 GB, is approximately 4 hours. In this example we have intentionally chosen the domain of $\mathcal{S}_2$ to be a quarter the size of $\mathcal{S}_1$, so that we have a virtual point source at $(-12,0,-20)$, which can clearly be seen in the figure. To verify the obtained reflectors, we ray-traced two-dimensional illuminance patterns and three-dimensional flux-density distributions, both with $10^6$ rays, where the targets are divided in $100\times100$ bins. The results are shown in Fig.~\ref{fig:ReflectorResults}, and clearly resemble the desired uniform square and circle intensity distributions.

\begin{figure}[!ht]
\centering
\begin{subfigure}[b]{0.22\textwidth}
    \centering
    \includegraphics[width=\textwidth,trim={0cm 0cm 0cm 0cm},clip]{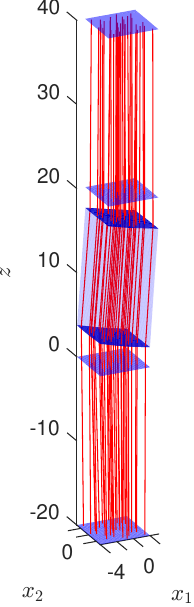}
    \caption{Lens system.}
    \label{fig:LensSystem}
\end{subfigure}
\hfill
\begin{subfigure}[b]{0.76\textwidth}
    \centering
    \begin{subfigure}[b]{0.48\textwidth}
        \centering
        \includegraphics[width=0.8\textwidth,trim={0cm 0cm 0cm 0cm},clip]{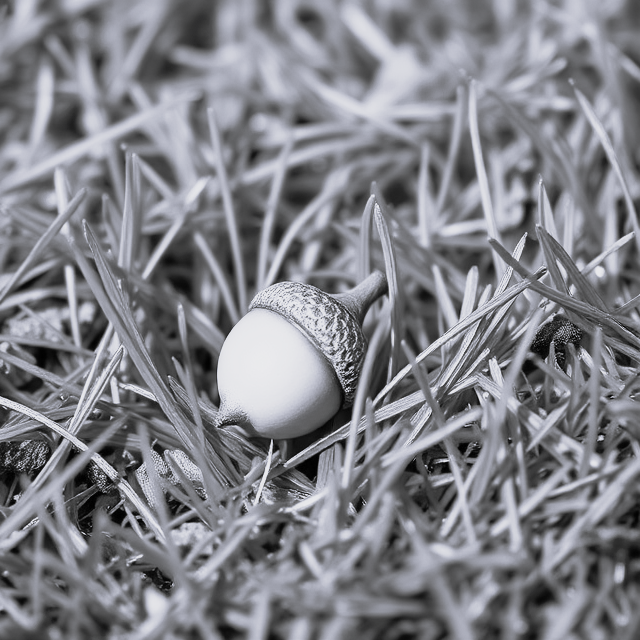}
        \caption{Distribution pattern of an acorn.}
        \label{fig:T1_image}
    \end{subfigure}
    \hfill
    \begin{subfigure}[b]{0.48\textwidth}
        \centering
        \includegraphics[width=0.8\textwidth,trim={0cm 0cm 0cm 0cm},clip]{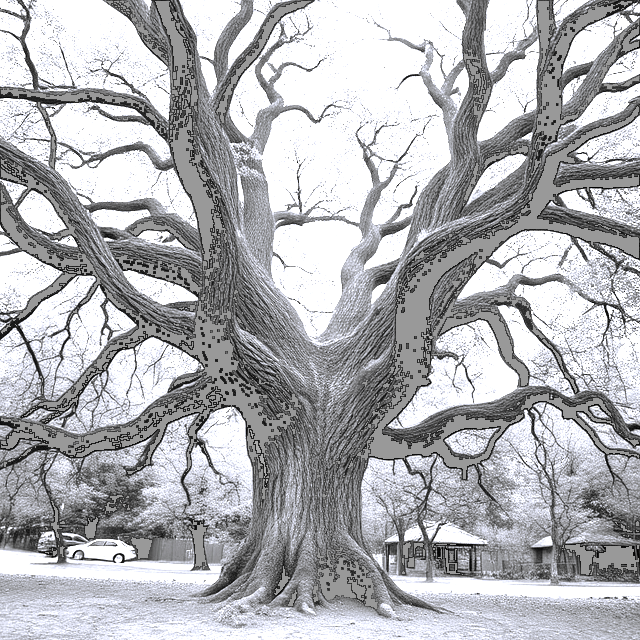}
        \caption{Distribution pattern of an oak tree.}
        \label{fig:T2_image}
    \end{subfigure}
    \begin{subfigure}[b]{0.48\textwidth}
        \centering
        \includegraphics[width=\textwidth,trim={0cm 0cm 0cm 0cm},clip]{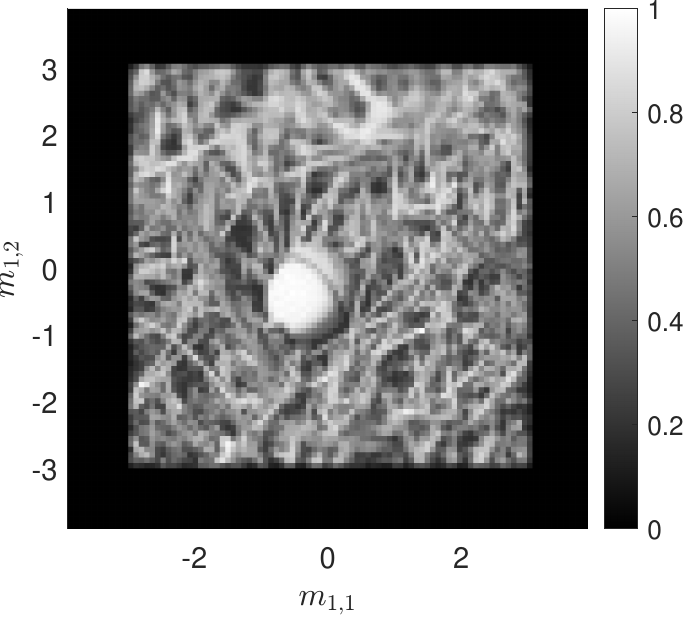}
        \caption{Ray-traced illuminance pattern on $\mathcal{T}_1$.}
        \label{fig:L_illuminance_pattern_T1}
    \end{subfigure}
    \hfill
    \begin{subfigure}[b]{0.48\textwidth}
        \centering
        \includegraphics[width=\textwidth,trim={0cm 0cm 0cm 0cm},clip]{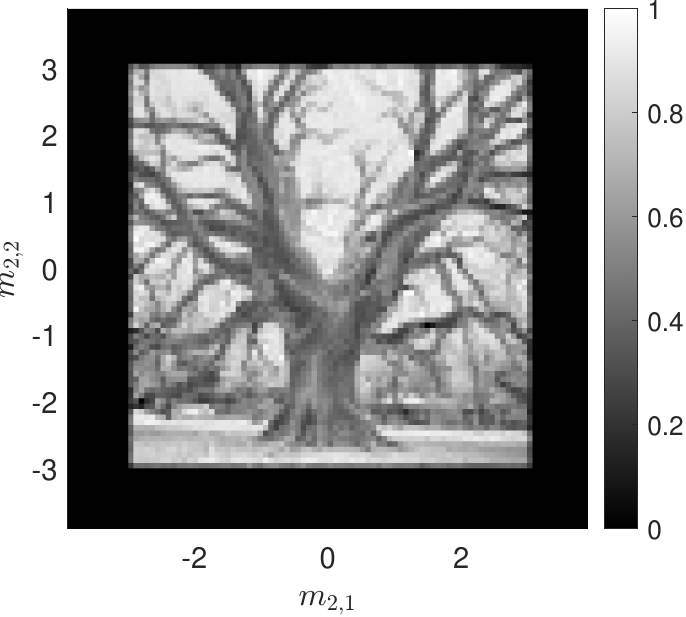}
        \caption{Ray-traced illuminance pattern on $\mathcal{T}_2$.}
        \label{fig:L_illuminance_pattern_T2}
    \end{subfigure}
\end{subfigure}
\caption{Results of the lens system with $n=1.5$, $L_0=-20$, $L_1=20$, $L_2=40$, $V_0=27$, $u_{1,0}=3$, computed after $100$ iterations on a $1000\times1000$ grid and verified by ray-tracing using $10^7$ rays and $100\times100$ bins.}
\label{fig:LensResults}
\end{figure}


\clearpage

\subsection{Lens system}
We consider the lens system with the domains $\bm{w}\in[-4,2]\times[-3,3]$, $\bm{x}\in[-4,2]\times[-3,3]$, $\bm{y}\in[-3,3]\times[-3,3]$ and $\bm{z}\in[-3,3]\times[-3,3]$. We set $n=1.5$, $L_0=-20$, $L_1=20$, $L_2=40$, $\alpha=0.5$, $V_0=27$, $u_{1,0}=3$ and take uniformly discretized grids of size $1000\times 1000$. We assume that the light emittance at $S_1$ is distributed according to a two-dimensional normal distribution with mean $(-1,0)$ and variance $2$, that the light emittance at $\mathcal{S}_2$ is distributed uniformly, that target intensity distributions at $\mathcal{T}_1$ and $\mathcal{T}_2$ correspond to distribution patterns of an acorn and an oak tree, as shown in Figs. \ref{fig:T1_image} and \ref{fig:T2_image}.

After $100$ iterations over each of the three stages in the algorithm, it finds the lens surfaces $\mathcal{L}_1$ and $\mathcal{L}_2$ illustrated in Fig. \ref{fig:LensSystem}. The computation time, on the same laptop as before, is approximately 8 hours. To verify these surfaces, we have computed ray-traced illuminance patterns on the two targets illustrated in Figs. \ref{fig:L_illuminance_pattern_T1} and \ref{fig:L_illuminance_pattern_T2} using $10^7$ rays, where the target is divided in $100\times100$ bins. They clearly display distribution patterns of the acorn and the oak tree, respectively. 

\section{Conclusion}\label{sec:conclusion}
This paper presents an inverse method to compute freeform optical surfaces that transform a general light source, parameterized by two planes, to two light distributions at different targets using either two freeform reflectors or two freeform lens surfaces. We used the optical path length to construct generating functions and used energy conservation to derive generated Jacobian equations. A three-stage least-squares algorithm was presented to compute the reflectors and lenses for complicated sources and targets, which we applied in two examples and which we verified by ray-tracing. These examples showed that the systems can handle point sources and far-field targets. Overall, the results advance our understanding of how the optical path length can be used to control both the position and the direction of a light ray at the target using an inverse method. In future work, we will include non-planar targets \cite{Jan2024_} and physical phenomena such as scattering effects \cite{Vi2023}. Moreover, since two target planes can serve as two new source planes of a subsequent identical system, we will also explore concatenation of optical systems.

\appendix

\begin{backmatter}
\bmsection{Funding}
This work was supported by the NWO Perspectief program \textit{Optical coherence; optimal delivery and positioning} (P21-20).

\bmsection{Disclosures}
The authors declare no conflicts of interest.

\bmsection{Data availability} Data underlying the results presented in this paper are not publicly available at this time but may be obtained from the authors upon reasonable request.
\end{backmatter}

\bibliography{Bibliography}

\end{document}